# Google Scholar: the 'big data' bibliographic tool

**Emilio Delgado López-Cózar[1], Enrique Orduna-Malea[2], Alberto Martín-Martín[1], Juan M. Ayllón[1]**

[1] Facultad de Comunicación y Documentación, Universidad de Granada, Colegio Máximo de Cartuja s/n, 18071, Granada, Spain.
[2] Institute of Design and Manufacturing, Universitat Politècnica de València, Camino de Vera s/n, 46022, Valencia, Spain.

* Corresponding author: E-mail address: edelgado@ugr.es.

**Abstract:** The launch of Google Scholar back in 2004 meant a revolution not only in the scientific information search market but also in research evaluation processes. Its dynamism, unparalleled coverage, and uncontrolled indexing make of Google Scholar an unusual product, especially when compared to traditional bibliographic databases. Conceived primarily as a discovery tool for academic information, it presents a number of limitations as a bibliometric tool. The main objective of this chapter is to show how Google Scholar operates and how its core database may be used for bibliometric purposes. To do this, the general features of the search engine (in terms of document typologies, disciplines, and coverage) are analysed. Lastly, several bibliometric tools based on Google Scholar data, both official (Google Scholar Metrics, Google Scholar Citations), and some developed by third parties (H Index Scholar, Publishers Scholar Metrics, Proceedings Scholar Metrics, Journal Scholar Metrics, Scholar Mirrors), as well as software to collect and process data from this source (Publish or Perish, Scholarometer) are introduced, aiming to illustrate the potential bibliometric uses of this source.

# 1. Introduction

Quantitative disciplines -such as Bibliometrics- are dependant to a great degree on their instruments of measurement. The more accurate the instrument, the better researchers will be able to observe specific phenomena. In the same way the telescope fostered the evolution of Astrophysics, improvements in bibliographic databases led to the advancement of Bibliometrics during the last decades of the 20$^{th}$ Century. The Internet (carrier), the Web (contents), and search engines (content seekers) did all play a role to change the paradigm of bibliographic databases. The coming of academic search engines was the beginning of the era of robometrics (Jacsó, 2011), where web-based tools automatically index academic contents, regardless of their typology and language, providing data to third-party applications that automatically generate bibliometric indicators. Google Scholar (GS) is the best robometric provider ever made. And GS never sleeps.

This chapter is devoted to the GS database. The main goal of this contribution is therefore to show how GS can be used for bibliometric purposes, and at the same time to introduce some bibliometric tools that have been built using data from this source.

# 2. Birth and development

GS is a freely accessible academic search engine (Ortega, 2014) which indexes scientific literature from a wide range of disciplines, document types, and languages, providing at the same time a set of supplementary services of great value. The fact that it displays the number of citations received by each document, regardless of their source, opened up the door to a new kind of bibliometric analysis, revolutionizing the evaluation of academic performance, especially in the Humanities and Social sciences (Orduna-Malea et al, 2016).

However, facilitating bibliometric analyses was never the main purpose of this platform. Google Scholar was conceived by two Google engineers (Anurag Acharya and Alex Verstak) who noticed that queries requesting academic-related material in Google shared similar patterns, and that these patterns were related to the structure of academic documents. Consequently, these requested contents might be filtered, offering a new specialized service aimed at discovering and providing access to online academic contents worldwide (Van Noorden, 2014). Bibliometric analyses were never a goal; this was just a byproduct brought on by its use of citations for ranking documents in searches.

Despite its simple interface and limited extra functionalities when compared to other traditional bibliographic databases, GS became rapidly known in the academic information search market after its launch in 2004. Both Science (Leslie, 2004) and Nature (Butler, 2004) reported the widespread use of this search engine among information professionals, scientists and science policymakers.

The evolution of its website (2004 to 2016) can be observed in Figure 5-1 through the yearly screenshots captured from the Internet archive's Wayback machine[1]. An austere homepage with a simple search box mimicking Google's general search engine and the "beta" declaration genuinely distinguished the first version. During these nearly twelve years since it was launched, the interface has barely changed. Google Scholar's improvements can't be clearly perceived from looking at its interface, because it's in the engine itself where changes have taken place.

**Figure 5-1. The evolution of Google Scholar (November 2004 to August 2016)**

**Source: Internet Archive**

Google Scholar's ease of use, simplicity, speed, as well as its multilingual and universal service, free of cost to the user, have contributed to its current popularity. To illustrate this, Figure 5-2 shows the worldwide search trends on Google for the main bibliographic databases (GS, Pubmed, Web of Science, and Scopus). Since science policies might differ by country, we also offer data from trends in particular countries (United States, Belgium, Colombia, and India). Complementarily, Figure 5-3 shows the popularity of the search terms in a sample of regions and cities.

**Figure 5-2. Bibliographic Databases Web Search Trends (I): Worldwide, United States, Belgium, Colombia and India**

Source: Google Trends

Note: red: "Pubmed"; blue: "Google Scholar"; yellow: "Web of Science"; green: "Scopus".

**Figure 5-3. Bibliographic Databases Web Search Trends (II): Regions and Cities**

Source: Google Trends

Note: red: "Pubmed"; blue: "Google Scholar"; yellow: "Web of Science"; green: "Scopus".

**3. Characteristics of the Google Scholar database**

The fundamental pillars that sustain GS's engine are generally unknown not only to the final users, but also to journal editors (who tend to forget that online journals are –at the end of the day- webpages) and information professionals (who still often think in terms of the classic bibliographic databases). The consequences of these misconceptions might constitute total web invisibility for publications that are not represented on the Web properly. Today, most students and researchers begin their searches of academic information in GS (Housewright et al, 2013; Bosman and Kramer, 2016). Thus, publications missing from Google Scholar's results pages may suffer significant losses in readership and maybe even a loss in citations as a result. This should be disquieting not only to journal editors and authors, however, but also to bibliometricians.

This section will describe how GS works as well as its main indexing requirements, with the hope that it will help users willing to extract -and contextualize- bibliometric data from this database.

*3.1. How does Google Scholar work? By capturing the academic web*

The approach of GS to document indexing clearly differs from classic bibliographic databases, which are based in the cumulative inclusion of selected sources based on their quality (mainly formal requirements). Google Scholar's approach, however, relies on the so-called "academic web": any seemingly academic document available online will be indexed as long as a series of technical requirements are met.

Fortunately for print-only journals, GS not only indexes individual contributions available online, but also online catalogues and directories. Therefore, all those contributions that are not available to Google Scholar's crawlers (because there isn't an online version, technical problems, or legal impediments like paywalls) but indirectly catalogued in other bibliographic products (such as Dialnet), will also be indexed in GS.

GS operates in a similar fashion to Google's general search engine, managing a net of automated bots that crawl the Web looking for relevant information. These web crawlers are trained to identify academic resources, extract their metadata and full texts (when available), and lastly to create a bibliographic record to be included into Google Scholar's general index. If Google Scholar's crawlers are able to access the full text (either directly or indirectly by agreements with publishers), the system will also analyse the cited references in the document, and these references will be linked to the corresponding bibliographic records in GS as citations. The entire process is completely automated.

In short, and overlooking for now some exceptions that will be discussed below, GS only indexes "academic resources" deposited in the "academic web" that meet certain web requirements. These intellectual and technical formalities are detailed below:

a) Academic web

The natural home for academic resources should be the academic web. This is Google Scholar's philosophy. Through the years they have created a list of diverse institutions -both public and private- related in some way to academia, such as higher education institutions, national research councils, repositories, commercial publishers, journal hosting services, bibliographic databases, and even other reputed academic search engines.

Once indexed, these places are regularly visited by GS's spiders. Aside from these well-known academic entities, any natural or legal person is allowed to request their inclusion in Google Scholar's academic web space through the GS inclusions service[2]. Some of the accepted Website types are: DSpace, EPrints, other repositories, Open Journal Systems (OJS), other journal websites, and personal publications. Table 5-1 shows the number of records indexed in GS extracted from a small assortment of academic entities and services. However, these data should only be considered rough estimations. This table provides a cursory view of the wide and diverse nature of the academic web space from which GS extracts information.

**Table 5-1. A Showcase of Academic Entities size in Google Scholar**

| TYPE | ENTITY | URL | RECORDS ALL | 2015 |
|---|---|---|---|---|
| Universities | *Harvard University* | harvard.edu | 2,260,000 | 53,000 |
| | *National Autonomous University of Mexico* | unam.mx | 80,900 | 4,870 |
| Research Organizations | *National Institute of Informatics* | nii.ac.jp | 12,900,000 | 279,000 |
| | *Max Planck Gesellschaft* | mpg.de | 105,000 | 4,240 |
| Thematic repositories | *Arxiv* | arxiv.org | 402,000 | 53,200 |
| | *Social Science Research Network* | ssrn.com | 380,000 | 37,600 |
| Publisher platforms | *Elsevier* | sciencedirect.com | 8,750,000 | 483,000 |
| | *Nature Publishing Group* | nature.com | 449,000 | 30,500 |
| Delivery services | *Ingenta Connect* | ingentaconnect.com | 658,000 | 34,000 |
| Databases | *Pubmed* | ncbi.nlm.nih.gov | 3,620,000 | 105,000 |
| | *Dialnet* | dialnet.unirioja.es | 2,830,000 | 101,000 |
| Academic Search engines | *CiteseerX* | citeseerx.ist.psu.edu | 1,020,000 | 13,200 |
| | *ResearchGate* | researchgate.net | 1,580,000 | 145,000 |

Note: records obtained through "site" search command, e.g: site:harvard.edu

While this method for finding out the number of records GS has indexed from each domain has its own advantages, it also has some important shortcomings when performing bibliometric analyses, among others:

- If data in Table 5-1 were collected again, results might be very different, maybe even lower! This is caused by the dynamic nature of the Web (Lawrence and Giles, 1999). If a document becomes unavailable for any reason, Google Scholar will delete its presence from the database during one of its regular updates. Google Scholar's index reflects the Web as it is at any given moment. Past is just past.

- The same document may be deposited in different places (journal website, institutional repository, personal webpage, etc.). Since the user experience wouldn't be improved by displaying the same document several times for the same query, Google Scholar groups together different versions of the same document (Verstak and Acharya, 2013). This process works fairly well for the most part, but it fails sometimes, mostly when the quality of the metadata is not very good, preventing the system from finding a match against its current document base for a new document it is about to index, when a different version of that document has in fact been indexed before. Since each "unclustered" version may receive citations independently, this problem affects any bibliometric analyses that might want to be carried out using Google Scholar data, and so it is indispensable to group together all existing versions manually before carrying out any citation analyses.

- The time elapsed since a resource becomes available online and Google Scholar's crawlers index it depends on the source. Harvard is Harvard, and indexing priorities do exist.

If we take a look at the last 1,000 articles published by the American Physical Society (APS)[3] indexed in GS (Figure 5-4), a 'step-like' indexing, unrelated to the official publication periodicity is observed. GS doesn't follow the classic and controlled issue-by-issue indexing process. Their technology made this practice obsolete, converting the academic web in a dynamic and uncontrolled web space. The irregularity and unpredictability of Google Scholar's indexing speed may bias some bibliometric analyses if it is not taken under consideration.

**Figure 5-4. APS Journals' indexation in Google Scholar (August 2016)**

b) Technical requirements

Google Scholar's inclusion policies also provide some guidelines for journal publishers and anyone who would like their contents to be correctly indexed in GS[4]. Some of these requirements are optional whereas others are compulsory. Failure to comply with these rules may provoke an incorrect indexing, or no indexing whatsoever for incompliant websites.

A first set of rules focuses on the websites. Their main objective is to ease content discovery. The website must not require users to install additional applications, to log-in, to click additional buttons, or use Flash, JavaScript, or form-based navigation to access the documents. In addition to that, the website should not display popups, interstitial ads or disclaimers.

A second set of rules is centered on the files that contain the full text. The system requires one URL per document (one intellectual work should not be divided into different files, while one URL should not contain independent works). Additionally, the size of the files must not exceed 5MB. Though larger documents will be described in GS, their full text (including cited references) will be excluded. This may bias bibliometric analyses since cited references included in doctoral theses and monographs (with files that tend to be larger than 5MB) will be omitted.

HTML and PDF are the recommended file types. Additionally, PDFs must follow two important rules: a) all PDF files must have searchable text. If these files consist of scanned images, the full texts (and cited references) will be excluded since Google Scholar's crawlers are unable to parse images; b) all URLs pointing to PDF files must end with the ".pdf" file extension.

Lastly, a third set of rules encourages resource metadata description, establishing some recommendations on compulsory fields (title, authors, and publication date), and preferred metadata schemes (Highwire Press, Eprints, BE Press and PRISM). Dublin Core may also be used as a last resort, but is not encouraged since its schema doesn't contain separate fields four journal title, volume, issue, and page numbers.

If no metadata is readily available in the HTML meta tags of the page describing the article, Google Scholar will try to extract the metadata by parsing the full text file itself. For this reason, GS also makes recommendations regarding the layout of the full texts: The title, authors, and abstract should be in the first page of the file (cover pages, used by some publishers, are strongly discouraged). The title should be the first content in the document and no other text should be displayed with a larger font-size. The authors should be listed below the title, with smaller font-size, but larger than the font-size use for the normal text. At the end of the document, there should be a separate section called "References" or "Bibliography", containing a list of numbered references.

### *3.2. The coverage of Google Scholar*

Coverage and growth rate are essential aspects of any bibliographic database. This is not only about information transparency but about context. Bibliometric analyses need to contextualize the results obtained since the database is just an instrument of measurement. Just like a chemist needs to check the calibration of his/her microscope to figure out the real dimensions of the observed elements in order to comprehend the underlying phenomena, an information scientist needs to verify which information sources are being indexed, the presence of languages, countries, journals, disciplines, and authors. Without context, Bibliometrics are just numbers. However, as the reader may probably imagine by reading the previous section, Google Scholar's coverage is, unfortunately, heterogeneous and still very much unknown.

Officially, GS indexes journal papers, conference papers, technical reports (or their drafts), doctoral and master's theses, preprints, post-prints, academic books, abstracts, as well as "other scholarly literature" from all broad areas of research. Patents and case laws are also included. Content such as news or magazine articles, book reviews and editorials is not appropriate for GS[4]. For example, any content successfully submitted to a repository will be included in GS regardless of its type. The lack of manual checking makes impossible to filter documents by document type.

Moreover, the absence of a master-list containing the publishers and sources that are officially covered has made many researchers wonder about its coverage. The continuous addition/removal of contents and sources as well as the technical exclusion of controlled sources make the elaboration of any master-list a chimera.

GS categorizes its documents into two independent collections: case laws and articles. The first group contains legal documents belonging to the Supreme and State courts of the United Stated. Since these documents are not used in Bibliometric analyses, we won't be studying them within this chapter. Regarding the collection of articles, we can distinguish the following contents:

a) Freely accessible online content

This group includes all resources for which Google Scholar is able to find a freely accessible full text link. If these documents include cited references, citing and cited documents will be automatically connected.

b) Subscription-based contents online

Most commercial publishers place the full-texts of the articles they publish behind paywalls. These documents are only accessible to people or institutions who have paid for the right to access them. By default, this would mean that Google Scholar might, at the most, have access to the basic bibliographic metadata for the articles (providing the publisher makes it available in the meta tags for each article, and doesn't block GS's spiders using robots.txt instructions), but probably not to the cited references, which are necessary to link citing and cited documents. Nowadays, however, GS has reached agreements with all the major publishers, and its spiders are able to collect all the necessary information from their websites (basic bibliographic metadata as well as cited references).

c) Content that is not available online

When Google Scholar's spiders parse cited references inside a document, the system checks for matches for those documents in its document base in order to build a citing/cited relationship. If a match is not found for any of these references (because it does not exist or any variation in the bibliographic description prevents a correct match), it is added to the document base as a citation record (marked as "[CITATION]" when they are displayed as a search result). There are two types of citation records (figure 5-5):

- Linked citations: some bibliographic references found by GS in library catalogues and databases, with no full text available.
- Unlinked citations: bibliographic references found in the "References" section of a full text already crawled.

**Figure 5-5. Linked and unlinked citations in Google Scholar**

d) Special collections

GS also indexes some collections from other Google services, such as Google Patents[5] and Google Book Search[6]. The inclusion guidelines are not very well documented in this regard, however.

GS officially states that the database automatically includes scholarly works from Google Book Search (excluding magazines, literature, essays, and the like). Additionally, any book cited by an indexed document will also be automatically included (as a [CITATION] record). In any case, users may also upload files directly to Google Books through their personal accounts. In a similar manner, the main reason to include patents is the fact that these resources are also cited in other documents indexed in GS. The fact that there are practically no patents in GS with 0 citations reinforces this assumption.

*3.3. The size of Google Scholar*

The growth of GS is dynamic and irregular as academic sources become "GS-compliant", new commercial agreements with publishers are attained, and old printed collections are digitized. This means that, apart from being continually indexing new materials as they are published, the retrospective growth of the platform is also remarkable. Figure 5-6 shows the number of records indexed in GS from 1700 to 2013 at two different times (May 2014 and August 2016). Although a logical correlation is obtained, differences are significant. For example, the difference between the two samples for the number of documents published in 2010 is of more than 1,770,000 records!

**Figure 5-6. Google Scholar evolution and retrospective growth**

**Note: number of documents extracted after sending queries using the "site" search command; books from: 'books.google.com'; Patents are not included: 'patents.google.com' and 'google.com/patents' does not work with the "site" command in GS.**

GS is updated several times a week. In addition to that, the whole database goes through a major update every 6-9 months, where it re-crawls the academic web and cleans data, making obsolete any study about its previous size. However, the matter of its size constitutes a hot topic for the scientific community, fuelled by the scarcity of official details on this issue. There has been several attempts to unveil the size of Google Scholar using a variety of methods, all described in the scientific literature (Jacsó, 2005; Aguillo, 2012; Khabsa and Giles, 2014; Ortega, 2014; Orduna-Malea et al, 2015).

Regarding the growth rate, Orduna-Malea et al (2015), De Winter, Zadpoor and Dodou (2014) and Harzing (2014) proved that GS grows faster than any traditional databases in all scientific fields.

However, even without considering the dynamic nature of the database, Google Scholar's size cannot be calculated accurately due to several limitations of the search interface: the custom time range filter is not accurate, the "site" command is not exhaustive, and the number of results per query is only a quick and rough estimation.

Savvy readers may be thinking of the possibility of performing several specific queries. However, no data export functionalities or API (Application Programming Interface) are available due to commercial constraints (probably some of the stipulations in the agreements with the publishers discussed above). Lastly, only the first 1,000 results of each query can be displayed. Some techniques like query splitting (Thelwall, 2008) may help but do not solve this limitation completely. Using web scraping in the results pages seems to be the only technical solution to perform big bibliometric analyses (see section 4.3), and even this approach has its own shortcomings. Moreover, doing this goes against Google Scholar's robots.txt, and they enforce this policy by blocking users who make too many queries too quickly.

## 4. Using Google Scholar for bibliometric purposes

While GS is a free, universal and fast search engine with an impressive coverage, it lacks key search functionalities for information professionals (advance filtering, export features, sorting options, etc.). It is oriented to final users and content discovery, and it is not designed to work as a bibliometric tool. These limitations may be what triggered the creation of applications which make use -either direct or indirectly- of its bibliometric data. We can distinguish between the official products designed by the

Google Scholar's team and other third party products created by external and independent research teams.

*4.1 The official products*

The GS team has designed two official products that make use of the bibliographic and bibliometric data available in the core database. One of them is focused on authors (Google Scholar Citations) and the other one on journals and the most cited articles in these journals (Google Scholar Metrics).

a) Google Scholar Citations (GSC)

Officially launched in November 16$^{th}$ 2011, this product lets users create an academic profile.[7] Users may build an academic résumé that includes all their contributions - providing they are indexed in GS. The publications will be displayed in the profile, sorted decreasingly by number of citations received by default (they can also be sorted by year of publication, and title). Users can search their publications using their own author name (with all its variants) or by searching documents directly in order to add them to the profile, merge versions of the same document that Google Scholar hasn't automatically detected, and fix bibliographic errors. Most importantly, profiles are updated automatically as GS indexes new documents, with the possibility of asking the author for confirmation before making any changes to the profile.

In addition to the number of citations, the platform provides 3 author-level metrics: total number of citations received, h-index, and i10-index (number of articles which have received at least 10 citations), which are available for all documents (useful to senior researchers) and for the documents published in the last 5 years (useful to emerging scholars).

The platform also offers additional services such as personalized alerts, lists of co-authors, areas of interests, list of authors by institution, etc. By using this product, users can improve the dissemination, and potentially, the impact of their contributions. In short, authors can track the impact of their papers other researchers' papers according to the data available in Google Scholar, as well as be constantly informed of new papers published by other authors. This makes Google Scholar Citations a very interesting, free, and easy to use research monitoring system.

The use of GS personal profiles can help to unveil much about an author's production and impact because, in a way, this product is a transition from an uncontrolled database to a structured system where authors, journals, organizations and areas of interest go through manual filters (Ortega, 2014). However, some of the problems in GS are also present in this product.

b) Google Scholar Metrics (GSM)

The Californian Company surprised the Bibliometrics community again in April $1^{st}$ 2012 by launching a journal ranking, commonly referred to as GS Metrics[8] (GSM). Its characteristics and functionalities make GSM a unique and original product: Since its inception, GSM presented important differences compared to other journal rankings like JCR and SJR (Cabezas-Clavijo & Delgado López-Cózar, 2012; Jacsó 2012). GSM is a bibliometric/bibliographic hybrid product, because in addition to bibliometric indicators, it also displays the list of most cited documents in each publication. Moreover, its selection policies, coverage (journals, repositories, and conferences), architecture, and formal presentation are also different to other journal rankings.

Its coverage in its first edition (publications with at least 100 articles published during the 2007-2011 period, and which had received at least 1 citation for those articles) and a categorization by language (for last version available, 2011-2015 period, it covers the following languages: Chinese, Portuguese, Spanish, German, Russian, French, Japanese, Korean, Polish, Ukrainian, and Indonesian; Dutch and Italian are deprecated) were two of its most distinctive features. However, displaying journals sorted by their h5-index (h-index for articles published in a given 5 year period) instead of using a similar formula to the widely criticized Journal Impact Factor or the impenetrable Scimago Journal Rank was probably its most distinctive feature. The platform also provided an internal search box which enabled users to locate journals not included in the general rankings, which were limited to the top 100 journals according to their h5-index. This search box presented the top 20 publications (again according to their h5-index) that matched the query terms entered by the user.

The lack of standardization, irreproducible data or the amalgam of publication typologies available in the first version (Delgado López-Cózar and Robinson-García, 2012) made Delgado López-Cózar and Cabezas-Clavijo (2012) consider GSM as an immature product, although acknowledging its potential as a source for evaluation of Humanities and Social Sciences journals in the future. The Google Scholar team fixed some of the deficiencies mentioned in those early reviews and launched an improved version in November 15[th] 2012, introducing a subject classification scheme composed of 8 broad categories and 313 subcategories. However, only journals published in English were classified in these categories, and only 20 publications were displayed in each of them (again the top 20 according to their h5-index).

Since then, the product has been updated every year. The last edition (July 2016) covers documents published in the 2011-2015 period. The total number of journals covered by this product is probably is over 40,000 (Delgado López-Cózar & Cabezas-Clavijo, 2013). At any rate, various studies confirm that GSM covers more journals, published in more languages and countries, than JCR and SJR (Repiso, Delgado López-Cózar, 2013; 2014; Reina, Repiso, Delgado López-Cózar, 2013; 2014; Ayllón et al. 2016).

Despite its continuous improvements, Martín-Martín et al. (2014) do not hesitate in labeling GSM as a "low cost" bibliometric tool, with some powerful advantages (coverage, simplicity, free of cost) but some important shortcomings, most of them related to the difficulties of processing of journal data automatically, without any human intervention.

## 4.2. Third party applications that collect and process Google Scholar data

### a) Harzing's Publish or Perish (PoP)

If any third party tool deserves a place in Google Scholar's Hall of Fame, this would undoubtedly be Publish or Perish (PoP). This free desktop application[9], officially launched in 2006, lets users send queries to Google Scholar and Google Scholar Citations (also to Microsoft Academic) and collect all the available data, edit it, and obtain a set of bibliometric indicators that can be exported outside the application. That is, all the features many GS users would like to be able to carry out natively from the official platform, and a few extras.

Despite its widespread use, the way PoP works is relatively unknown. Some people think it is an independent database, unrelated to Google Scholar, and some think it makes use of some special API to access the information available in GS. None of that is true. Publish or Perish serves as a friendly interface between the user and GS (Harzing, 2013), but it is subjected to the same limitations of any normal search in GS. The application transforms a user query, and makes the appropriate request directly to Google Scholar's advance search. It then parses the results, displaying them on the PoP interface, at the same that it calculates additional metrics (total number of citations, authors per article, h-index, g-index, e-index, generalized h-index, AR-index, hlnorm, hl annual, multiauthored h-index, etc.).

Without any doubt, Google Scholar's coverage, together with Publish or Perish, has contributed to the democratization and popularization of citation analyses (Harzing and Van der Wal, 2008).

*b) Scholarometer*

This one is a less known but powerful tool developed by the School of Informatics and Computing at Indiana University-Bloomington, launched in 2009 (Kaur et al, 2012). It is a social tool which intends not only to facilitate citation analysis but also to facilitate social tagging of academic resources.[10]

Scholarometer is installed as a web browser's extension and focuses primarily on the extraction of author data from GSC. Users can search authors through a search box or alternatively introduce the scholar's ID. In this last case, the system will extract and display the author's GSC profile, adding some extra functionalities and metrics, such as the article rank, and providing some data export functionalities as well, which are not available natively in Google Scholar Citations (except for one's own profile).

*4.3. Third party products that make use of Google Scholar data*

Lastly, we'll introduce a set of bibliometric products designed and developed by the EC3 Research Group in Spain. The purpose of this is to illustrate the sort of products that can be generated using data from GS, GSC, and GSM.

*a) H Index Scholar*

Bibliometric index which seeks to measure the academic performance of researchers from public Spanish universities in the areas of Social Sciences and Humanities (SS&H) by counting the number of publications and citations received by their publications, according to data from GS (Delgado López-Cózar et al, 2014). Rankings are displayed broken down by 4 broad areas of knowledge (Social Sciences; Humanities, Law, and Fine Arts), grouped into 19 disciplines and 88 fields[11]. In each field, authors were sorted according to their h-index and g-index, facilitating the identification of the most influential Spanish authors in the Humanities and Social Sciences as of 2012.

This project (which covered more than 40,000 Spanish researchers) was the first serious attempt to study the suitability of GS for collecting the academic output of all SS&H researchers in a country.

*b) Publishers Scholar Metrics*

Book citation data is fundamental not only for researcher-level assessment in the SS&H, but also to get an idea of the average impact made by each publisher. To date, there is no citation database that has anything resembling a comprehensive coverage of scientific books (the Book Citation Index is currently far from achieving this goal), and the evaluation of publishers has often been relied on reputational surveys.

Publishers Scholar Metrics[12] makes use of the GS database to construct a bibliometric index to find evidence of the impact of book publishers, based on the citations received by the books published by the authors indexed in H Index Scholar (researchers from public Spanish universities, working in the SS&H, data collected in 2012). The main objectives of this product were to identify the core publishers by field as well as demonstrating the suitability of GS data to carry out this task.

*c) Proceedings Scholar Metrics*

This product is a ranking of scientific meeting proceedings (conferences, workshops, etc.), in the areas of Computer Sciences, Electric and Electronic Engineering, and

Communications, which have been indexed in GSM. To date, two editions (2009-2013 and 2010-2014) have been published.[13-14]

The main objective of this ranking is to compile an inventory of all conference proceedings included in GSM in these fields of knowledge, ranking them by their h5-index according to data from GSM.

*d) Journal Scholar Metrics (JSM)*

This bibliometric tool seeks to measure the performance of Art, Humanities, and Social Science journals by means of counting the number of citations their articles have received, according to the data available in GSM.[15]

The main goals of JSM are, firstly, to calibrate the degree to which GSM covers international journals in the Arts, Humanities and Social Sciences, and secondly, to identify the core journals in each of the disciplines (as well as other related journals with thinner ties to the discipline) while offering a battery of citation-based metrics, complementing the indicators provided natively by GSM. By processing the metadata available in GSM, all journal self-citations (also called self-references) have been identified, allowing the calculation of the number of citations for each article excluding these journal self-citations, and consequently, the h5-index excluding journal self-citations and the journal self-citation rate.

*e) Scholar Mirrors*

This last product is a multi-faceted platform which aims to quantify the academic impact of a whole scientific community (as a case study, we analyzed the community of researchers working on Scientometrics, Informetrics, Webometrics, and Altmetrics) [16].

The development of this product started with the identification of all the key members belonging to the studied community. To do this, those authors in the field who had created a public GSC profile were identified, and all the bibliographic information related to each of their contributions was collected.

Scholar Mirrors could be considered a deconstruction of traditional journal and author rankings, in alignment with the notion of multi-level analyses of a scientific discipline (documents, authors, publishers, and specific topics) instead of evaluating authors exclusively through the impact reported in the communication channels in which the research findings are published.

The platform offers a battery of author-level indicators extracted from a wide variety of academic social networks and profile services, making a total of 28 indicators (6 from Google Scholar Citations, 5 from ResearcherID; 9 from ResearchGate; 4 from Mendeley; 4 from Twitter). Additionally, authors are categorized as core (those authors whose scientific production is primarily done in the field of Bibliometrics) and related (those authors who have sporadically published bibliometric studies, but their main research lines lie in other fields). The elements in the remaining levels of analysis (documents, journals, and publishers) are ranked according to the aggregated number of citations.

This multi-faceted model facilitates the observation of the performance of the elements at different levels at the same time. It also displays in a clear way what each platform reflects about each of the authors, by way of their respective indicators, hence the "mirror" metaphor.

## 5. Final remarks

Before presenting some conclusions on the potential bibliometric uses of GS, we should stress an essential point which was already mentioned at the beginning of the chapter: the underlying reason that explains much of what has been discussed here. GS is, first and foremost, an academic search engine, a gateway to finding scientific information in the Web. It was conceived with this sole purpose, and all its features and improvements are oriented to further this goal: connecting researchers with studies that may be useful to them. It was the members of the bibliometric community who, upon becoming aware of the sheer wealth of scientific information available within this search engine, have repeatedly insisted on using this platform as a source of data for scientific evaluation. Once we acknowledge this truth, it is easy to understand the limitations of this platform for bibliometric analyses.

At this point, curious readers may be asking themselves a fundamental question. Given the special nature of GS (unassisted, uncontrolled, unsupervised, lacking many advanced search features)… not to mention all its errors (authorship attribution, false citations, gaming, etc.), impossible to enunciate and describe in detail in a single book chapter… Why use GS? The answer revolves around a concept: big data.

GS is undoubtedly the academic database with the widest coverage at this time, including journal articles, books, book chapters, dissertation theses, reports, proceedings, etc. No other scientific information system covers as many document types as GS, representing practically all formal and informal academic dissemination practices. Crawling the academic web allows GS to collect a great percentage of citations that are undetectable to other classic citation databases: GS is an academic big data system.

Classic citation databases constitute closed environments, and are based on the controlled selection of a discrete number of journals: the elite. Classic citation indexes are built on citations among articles published in elite journals. This limitation - sometimes passed off as a feature- has its roots in Bradford's law on literature concentration: few journals control a great percentage of science advancements. In practice, applying this law to select only a small portion of the scientific literature was primarily done because of technological and economic constraints which have now disappeared for the most part.

When GS started collecting data about all seemingly academic documents deposited in trusted web domains (but which had not necessarily passed any other external controls, like peer review), they broke with those, until then, common selection practices, to which researchers (as well as journal editors) had gotten used to. GS didn't limit itself to the scientific world in the strict sense of the term, but instead embraced the whole academic world.

Although important errors do exist, big data transforms them in inherent aspects of the database. Even with "dirty" data, it is able to distinguish the wheat from the chaff. However, citations mixed regardless the source represent the real fraught between apocalyptic and integrated. Citations among any academic resource conforms the Google Scholar's author credential. The excellence view is short-sighted.

Citation-based author performance evaluation through traditional bibliographic databases (WoS or Scopus) for researchers in the Humanities and Social Sciences makes no sense, as these platforms do not cover the main venues where these authors disseminate their research results. The launch of Google Scholar back in 2004 meant a revolution not only in the scientific information search market but also for research evaluation processes, especially for disciplines where results are not usually published as articles in journals published in English. In order to cover the need to evaluate those disciplines, several applications and products that make use of Google Scholar data (with a much better coverage of the research outputs in those disciplines) have been developed. Google Scholar presents many challenges, but also a lot of potential as a source of data for bibliometric analyses.

**Notes**

1. https://archive.org/web

2. https://partnerdash.google.com/partnerdash/d/scholarinclusions

3. http://journals.aps.org

4. https://scholar.google.com/intl/en/scholar/about.html

5. https://patents.google.com

6. https://books.google.com

7. https://scholar.google.com/citations

8. https://scholar.google.com/metrics

9. http://www.harzing.com/pop.htm

10. http://scholarometer.indiana.edu

11. http://hindexscholar.com

12. http://www.publishers-scholarmetrics.info

13. http://arxiv.org/abs/1412.7633

14. http://doi.org/10.13140/RG.2.1.4504.9681

15. http://www.journal-scholar-metrics.infoec3.es

16. http://www.scholar-mirrors.infoec3.es

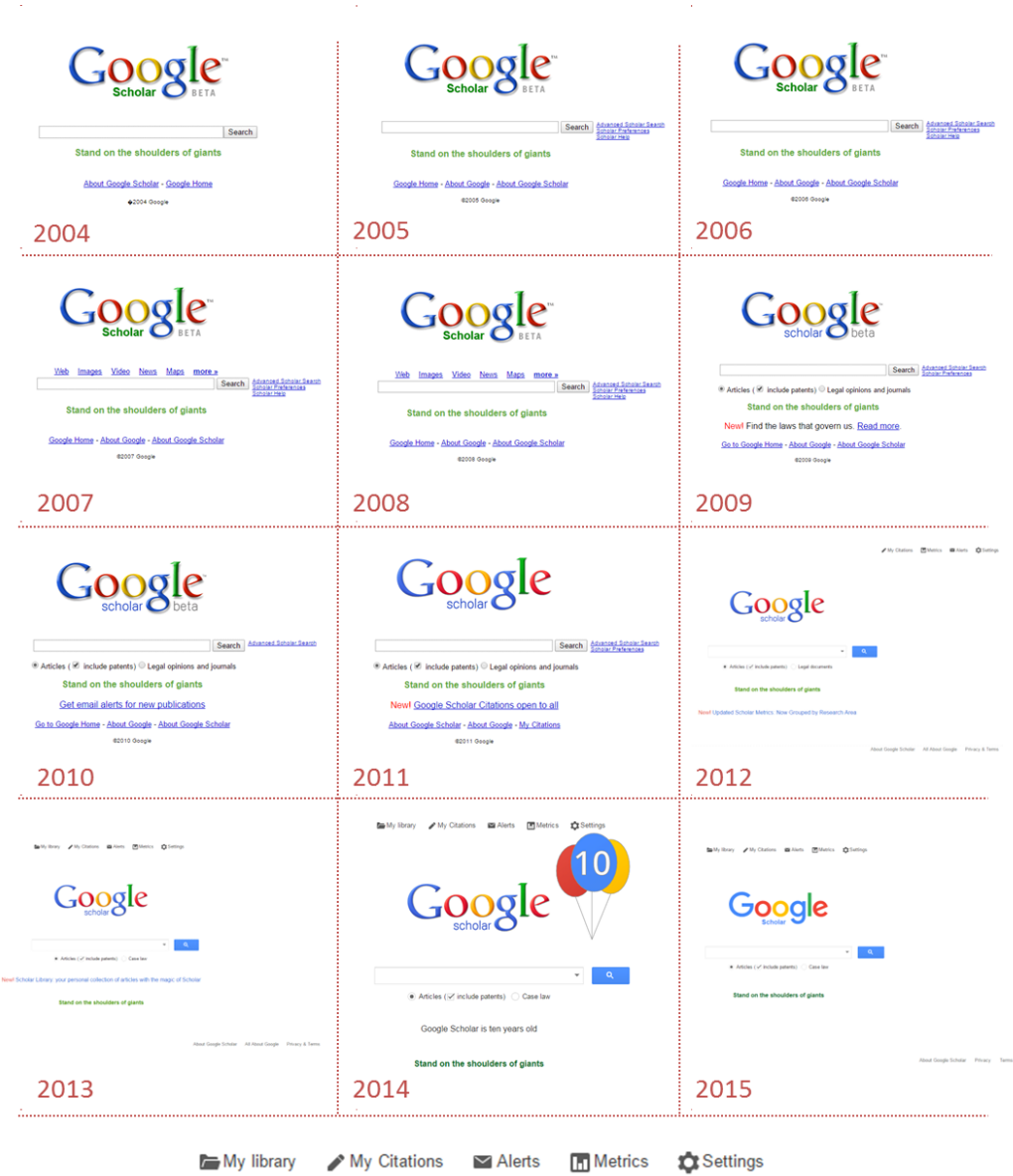

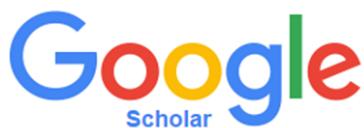

2016

**Figure 5-1. The evolution of Google Scholar (November 2004 to August 2016)**

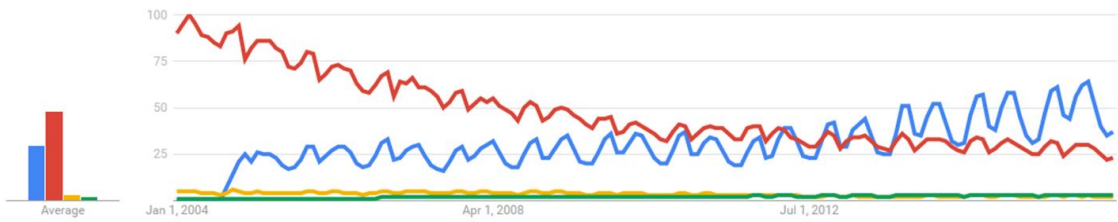

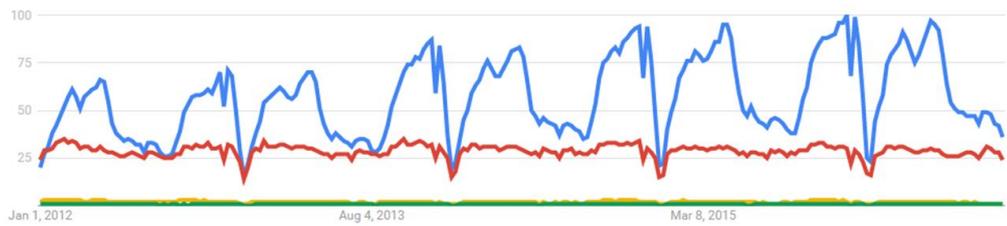

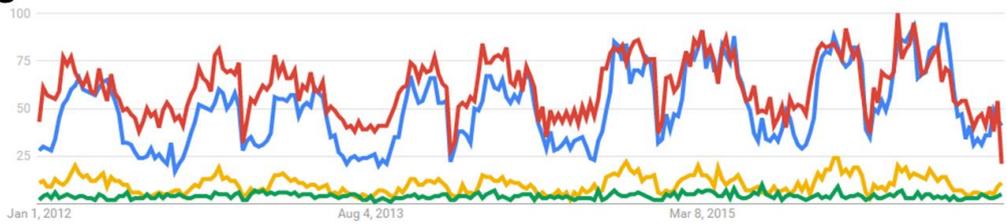

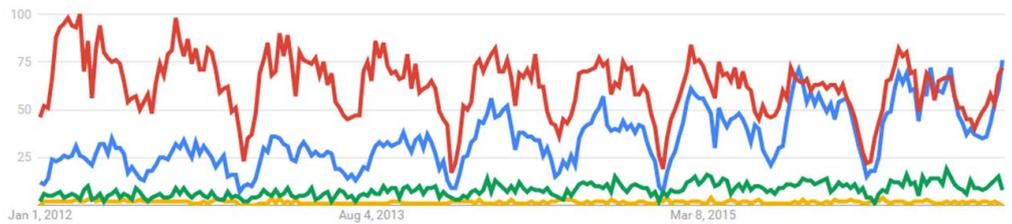

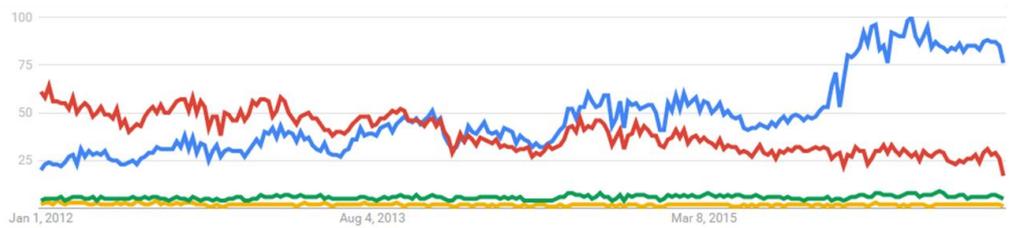

**Figure 5-2. Bibliographic Databases Web Search Trends (I): Worldwide, United States, Belgium, Colombia and India**

## Regions

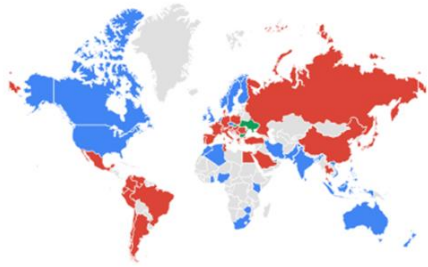
**All period**

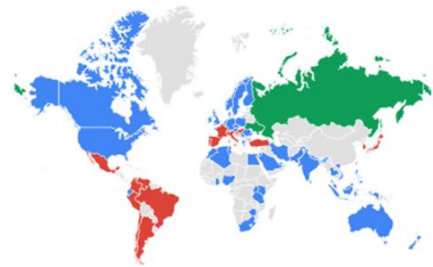
**Last year**

● Pubmed  ● Google Scholar  ● Web of Science  ● Scopus

## Cities

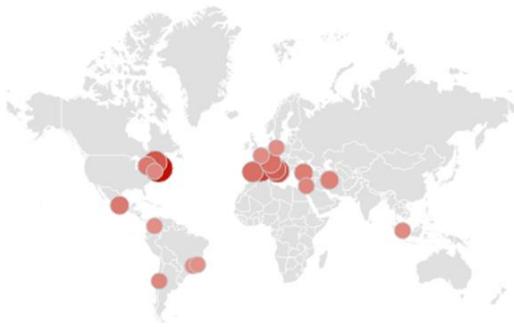
**Pubmed**

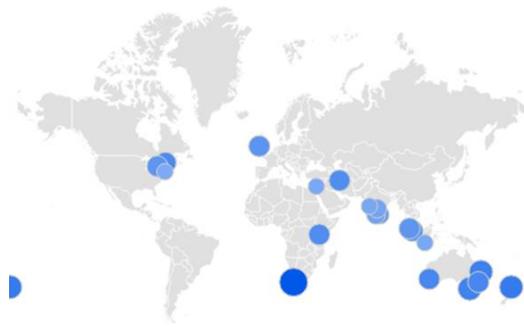
**Google Scholar**

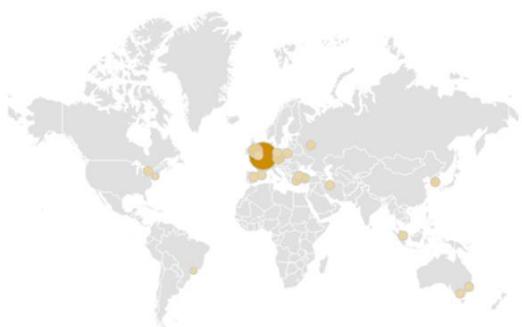
**Web of Science**

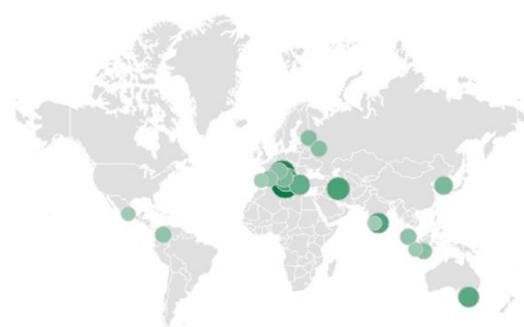
**Scopus**

All period: from January 2004 to July 2016
Last year: from August 2015 to July 2016

**Figure 5-3. Bibliographic Databases Web Search Trends (II): Regions and Cities**

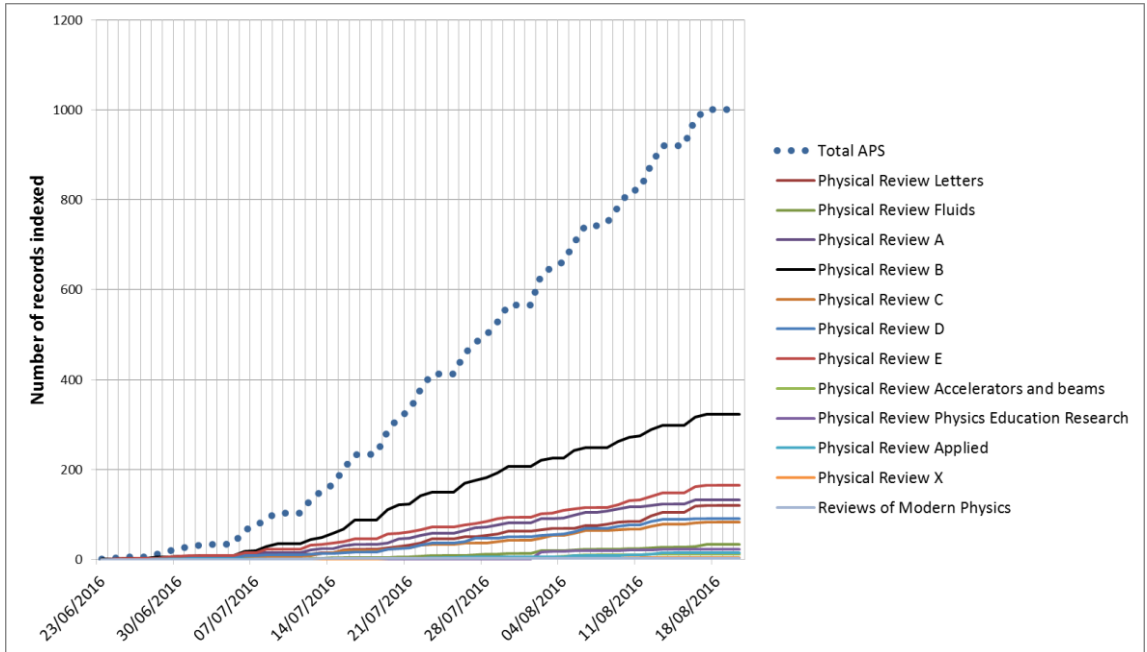
**Figure 5-4. APS Journals' indexation in Google Scholar (August 2016)**

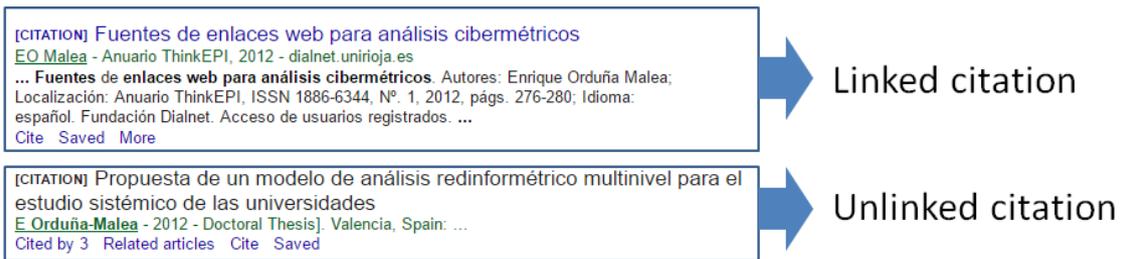
**Figure 5-5. Linked and unlinked citations in Google Scholar**

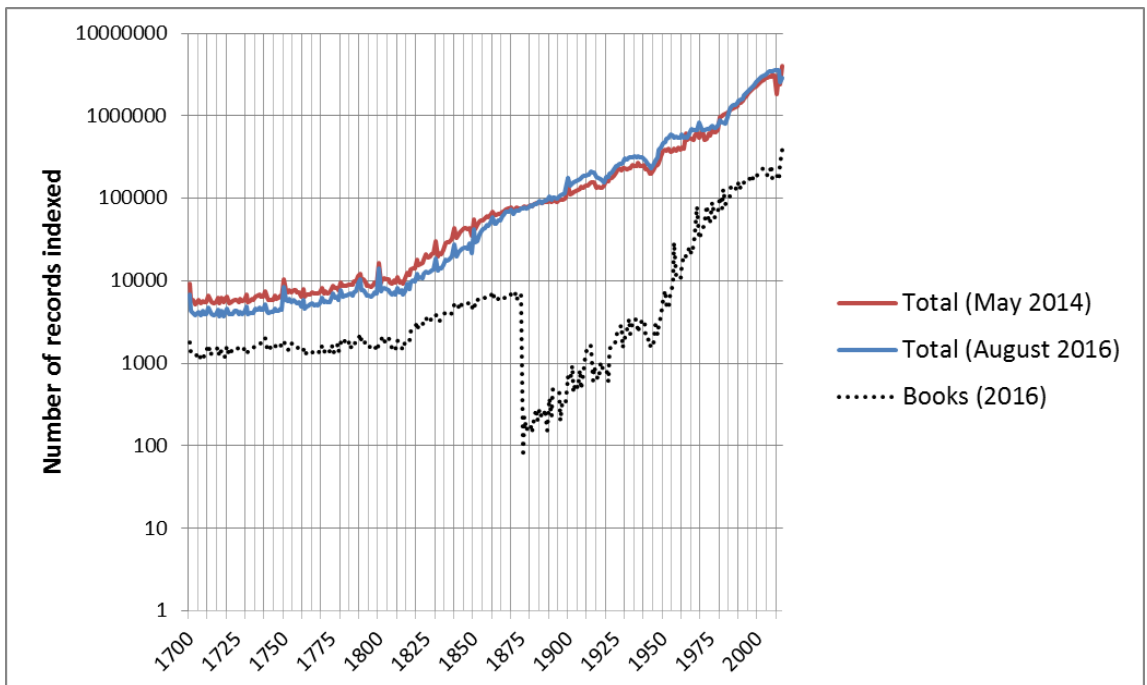
**Figure 5-6. Google Scholar evolution and retrospective growth**